# NEW GIANT RADIO GALAXIES IN THE SDSS-JVLA STRIPE82 AND LOTSS-PDR SURVEYS

Jonatan Rentería Macario[1] and Heinz Andernach[2]

## RESUMEN

Las radiofuentes extragalácticas con tamaño mayor que un Megaparcec (1 Mpc = 3.09 x $10^{22}$ m = 3.3 x $10^{6}$ años luz) se llaman Radiogalaxias Gigantes (GRG). En los últimos años se ha podido cuadruplicar el número de GRGs publicadas en la literatura mediante una inspección visual de radiorastreos de gran escala como el *NRAO VLA Sky Survey* (NVSS) y *Faint Images of the Radio Sky at Twenty Centimeters* (FIRST). Aquí reportamos el descubrimiento de 7 GRGs previamente desconocidas en dos rastreos recientes, el *JVLA 1-2 GHz Snapshot Survey of SDSS Stripe82* y *LOFAR Two-meter Sky Survey Preliminary Data Release* (LoTSS-PDR) a 150 MHz.

## ABSTRACT

Extragalactic radio sources with projected linear size larger than one Megaparsec (1 Mpc = 3.09 x $10^{22}$ m = 3.3 x $10^{6}$ light years) are called Giant Radio Galaxies (GRGs) or quasars (GRQs). Over the past few years our search for such objects by visual inspection of large-scale radio surveys like the *NRAO VLA Sky Survey* (NVSS) and *Faint Images of the Radio Sky at Twenty Centimeters* (FIRST) has allowed us to quadruple the number of GRGs published in literature. Here we report the discovery of 7 new GRGs in two recent surveys, the *JVLA 1-2 GHz Snapshot Survey of SDSS Stripe82* and the 150-MHz *LOFAR Two-metre Sky Survey Preliminary Data Release* (LoTSS-PDR).

**Palabras Clave:** rastreos en radio, radiofuentes extragalácticas, radiogalaxias.

## INTRODUCTION

Classical double radio galaxies have a median projected linear size (LLS) of between 150 and 200 kpc, and many thousands of them are known. Only a small fraction of these exceed an LLS of 1 Mpc (we use $H_0$ = 70 km $s^{-1}$ $Mpc^{-1}$, $\Omega_m$=0.3, $\Omega_\Lambda$=0.7 throughout). The largest known GRG and GRQ are J1420−0545 with 4.8 Mpc (Machalski et al. 2008) and J0931+3204 with 4.3 Mpc (Coziol et al. 2017). Since 2012 one of us (H.A.), aided by various students, used visual inspection of large-scale radio surveys like NVSS (Condon et al. 1998) and FIRST (Helfand et al. 2015) to quadruple the number of known GRGs from ~250 reported in literature to over 1000 (Andernach et al. 2012, Santiago-Bautista et al. 2016). The extreme sizes of GRGs were initially attributed to (a) a preferred orientation in the sky plane, (b) a location in low-density environments, or (c) more powerful jets, but neither of these were consistent with observational evidence (Komberg & Pashchenko 2009). In fact, applying our own algorithm (Ortega-Minakata et al. 2013) to a sample of GRGs with redshift z<0.3 to count the number of neighbor galaxies, we found no relation between the LLS and the number of neighbor galaxies within 1 Mpc from the GRG host galaxy. However, due to their huge extent GRGs serve as tracers of the large-scale galaxy distribution (Malarecki et al. 2015). Since late 2013 the volunteers of the Radio Galaxy Zoo citizen science project (Banfield et al. 2015) found over 200 new GRGs. A preliminary review of 1000 GRGs collected by one of us until December 2016 (Andernach 2016) showed several unexpected results which motivated us to look for further GRGs in more recent surveys, which will also provide us with a better estimate of the potential GRG content of planned next-generation surveys.

## METHODS AND MATERIALS

For our search we selected two recent radio surveys that cover large areas of sky. The first of these (Heywood et al. 2016, referred to in what follows as HJB2016) was made with the *Jansky Very Large Array* (JVLA) in NM, USA, covering 100 $deg^2$ at 1-2 GHz with an angular resolution of 16"x10". Observed in CnB configuration, it is sensitive to structures extending up to ~16'. Another advantage for the identification of radio galaxies is that the area is part of the *Sloan Digital Sky Survey* (SDSS) *Stripe82* along the celestial equator, which has an exceptionally deep optical coverage (Annis et al. 2014), and was previously observed with the VLA at 1.8" resolution by Hodge et al. (2011).
The second survey is LoTSS-PDR (Shimwell et al. 2017) and was made with LOFAR, a novel radio interferometer in the Netherlands and neighboring countries. It covers 350 $deg^2$ at low frequency (120-168 MHz) and angular resolution of 25", between right ascension $10^h30^m$ and $15^h30^m$ and declinations +45º and +57º. Thanks to LOFAR's shortest baseline of <50 m, LoTSS-PDR has an unprecedented sensitivity to extended emission on degree scales.

[1]Universidad Autónoma de Zacatecas, Calzada Solidaridad esq. Paseo La Bufa s/n, 98060, Zacatecas, Zac. México  jonatan.renteria.m@gmail.com
[2]Departamento de Astronomía, DCNE, Universidad de Guanajuato, Guanajuato, México   heinz@astro.ugto.mx

In order to search for extended features in these surveys, which would suggest the presence of large radio galaxies, we displayed the images with the program `ds9` (see http://ds9.si.edu) together with markers at the positions of radio galaxies already known from the compilation of one of us (Andernach 2016) and thus prevent their rediscovery. New candidates were marked with a circle enclosing their total radio extent and saved as a "region file". After cleaning the resulting source lists from duplicates in the overlap regions of neighboring images, a total of 177 candidates from HJB2016 and 2055 from LoTSS-PDR remained, with ~25 % high-, 25 % medium-, and 50 % low-priority objects, depending on the shape and/or presence of radio bridges between candidate lobes. To assess the reality of these and find optical identifications, we used radio images from NVSS, FIRST, VLAS82 (Hodge et al. 2011), and TGSS-ADR1 (Intema et al. 2017) as well as optical images and redshifts from SDSS DR14 (Abolfathi et al. 2017). We used photometric redshifts ($z_{ph}$) from various sources (cf. Andernach 2016) to estimate the distance of the GRG hosts, assuming $z_{ph}$>0.5 if undetected in SDSS. Only high-priority objects were inspected in the five weeks of this project.

For a few objects we used the `Aladin` software (Bonnarel et al. 2000) to measure geometrical parameters like the angular distance (or "armlength") from the host galaxy to the outer extremes of the radio lobes on each side, the orientation of each arm on the sky, and to integrate the flux of extended emission regions. We thus determined the armlength ratio (ALR) as the ratio of stronger-to-fainter lobe length, the bending (or misalignment) angle (BA) between the arm orientations, as well as the flux ratio (FLR) in the sense of longer-to-shorter arm integrated flux.

## RESULTS

Table 1 lists the seven GRGs newly found by us in the two surveys. Columns are (1) the survey (a for HJB2016, b for LoTSS-PDR), (2) the short GRG Jname used in this text, (3) the host galaxy name, (4) the largest angular size in arcmin, (5,6) the redshift (`s' if spectroscopic and `p' if photometric), (7) the largest projected linear size in Mpc, and (8) the host's magnitude and the corresponding filter. All radio images in this paper have north up and east to the left.

*Table 1. Giant Radio Galaxies discovered in HJB2016 (a) and LoTSS-PDR (b)*

| (1) | (2) | (3) | (4) | (5) | (6) | (7) | (8) | |
|---|---|---|---|---|---|---|---|---|
| a | J0152+0015 | SDSS J015214.39+001502.1 | 2.24 | 0.8438 | s | 1.03 | 23.34 | r' |
| a | J2245-0032 | SDSS J224520.76-003206.1 | 2.98 | 0.66 | p | 1.25 | 21.47 | r' |
| b | J1058+5140 | SDSS J105817.89+514017.7 | 4.5 | 0.415 | s | 1.51 | 19.65 | r' |
| b | J1102+5257 | PSO J110245.010+525757.76 | 3.57 | > 0.5 | p | >1.31 | 20.5 | i |
| b | J1139+4726 | AllWISE J113912.66+472657.8 ? | 3.8 | > 0.5 | p | >1.39 | 20.5 | i |
| b | J1139+4721 | AllWISE J113930.78+472153.4 | 4.25 | > 0.5 | p | >1.56 | 17.73 | W2 |
| b | J1301+5105 | SDSS J130125.90+510500.6 | 10.3 | 0.275 | p | 2.59 | 18.60 | r' |

Two new giants were found in HJB2016. Figure 1 shows the NVSS contour map of GRG J0152+0015 on the left, evidencing an extreme flux ratio of FLR=0.043 for a moderate ALR=0.61 and BA=7°. The top right panel shows the same source in the HJB2016 survey, the red cross marking the host galaxy with its radio nucleus. The bottom right image at 1.8" resolution is from Hodge et al. (2011) and shows the trail from the SW hotspot towards the nucleus, while the NE hotspot remains almost unresolved.

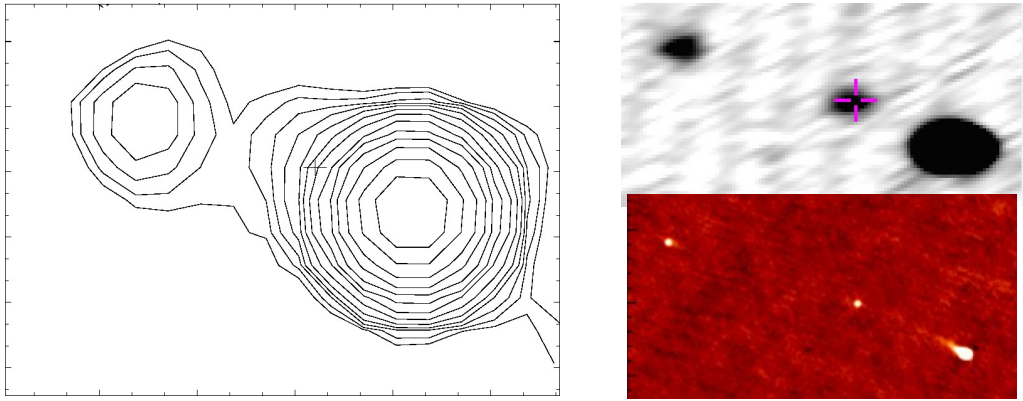

*Figure 1. GRG J0152+0015 in NVSS (left), HJB2016 (top right) and Hodge et al. 2011 (lower right). See text.*

Figure 2 shows the HJB2016 image of GRG J2245−0032 on the left, the FIRST image in the middle, and the high-resolution image from Hodge et al. (2011) on the right. In each panel the gapped cross marks the host galaxy with a radio core only barely detected at ~0.3 mJy in HJB2016, corresponding to a radio luminosity of log $P_{1.4GHz}$ ~ 23.5 W/Hz. Its ALR, FLR and BA are 0.88, 0.6, and ~10º, respectively. In NVSS only two unconnected extended sources (the lobes) appear aligned with each other, with a total flux of ~16 mJy, implying a total log $P_{1.4GHz}$ ~ 25.3 W/Hz.

In LoTSS-PDR five new GRGs were discovered. Figure 3 shows LoTSS-PDR and FIRST images of J1102+5257. It has an ALR of 0.69 and from NVSS we obtain a flux ratio of FLR=0.25. and from FIRST a very small BA of <2°. Note that the median BA for 200 GRGs was found to be ~5° by Jiménez Andrade (2015). The host is undetected in SDSS but shows up in the near infrared in PanSTARRS (Flewelling et al. 2016) and in the mid-infrared in AllWISE (Cutri et al. 2013). Assuming a redshift $z_{ph}$>0.5 its angular size of 3.57' implies an LLS of >1.3 Mpc; its integrated fluxes at 150 MHz and 1.4 GHz thus yield values of log$P_{150MHz}$ > 26.3 W/Hz and log$P_{1.4GHz}$ > 25.3 W/Hz.

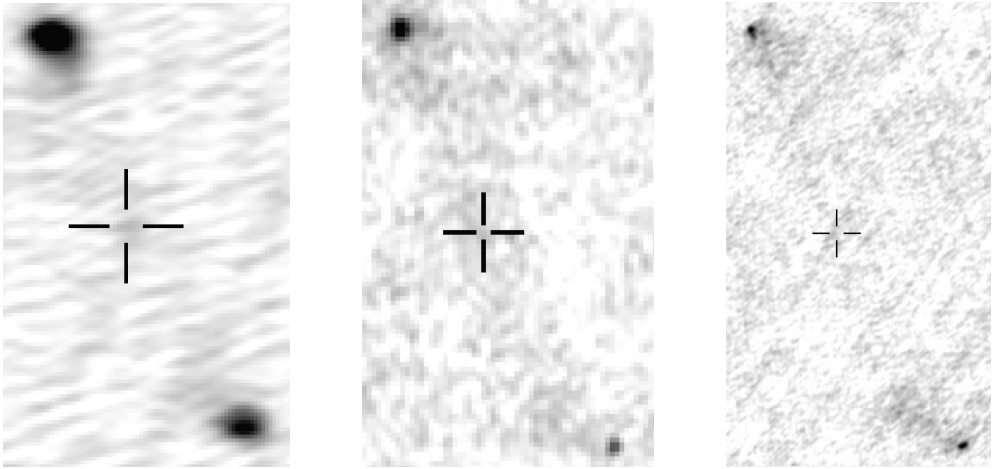

*Figure 2. 1.4-GHz images of GRG J2245-0032. Left: HJB2016, middle: FIRST, right: Hodge et al. (2011).*

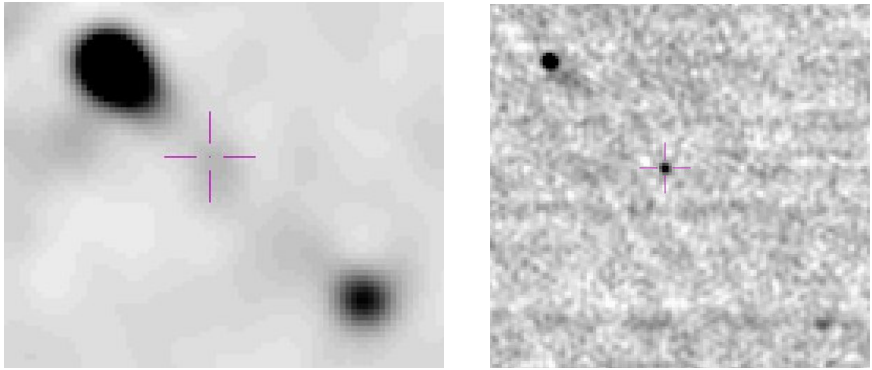

*Figure 3. GRG J1102+5257 in LoTSS-PDR at 150 MHz (left) and in FIRST at 1.4 GHz (right). See text for details*

Our most unusual finding is the pair of parallel GRGs J1139+4721 and J1139+4726 with sizes of 4.25' and 3.8' and oriented at position angles of 149° and 147°, respectively, with their most likely hosts separated by 6.1' along position angle 149°. Neither host galaxy is detected in SDSS nor PanSTARRS, suggesting $z_{ph}$ > 0.5 for both. AllWISE reveals both of them to have mid-infrared colors typical of Active Galactic Nuclei (AGN): W12,W23=+0.19,+3.91 for J1139+4726 and W12,W23,W34=+0.54,+4.66,+3.4 for J1139+4721. Another possible host for the latter GRG, SDSS J113931.77+472124.3 =AllWISE J113931.77+472124.4, $z_{sp}$=0.518, would not change this situation, implying a host separation of 2.2 Mpc on the sky with each GRG having sizes of 1.6 and 1.4 Mpc. In Figure 4 the NVSS image shows just a chain of 5 sources, but the overlay of LoTSS-PDR contours on the FIRST image in grey shows that each radio galaxy has diffuse lobes at opposite ends, pointing towards a central object with faint FIRST emission and coinciding with mid-infrared objects in AllWISE (red crosses in the left panel, red diamonds on the right). Big and small blue circles at bottom right mark the beam sizes of LoTSS-PDR and FIRST.

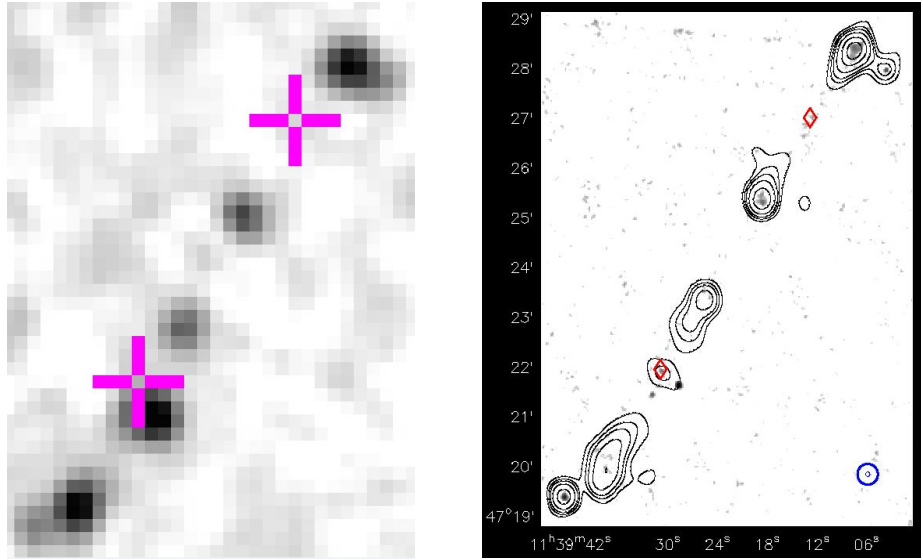

*Figure 4. GRGs J1139+4721 and J1139+4726. Left: NVSS. Right: LoTSS-PDR contours on FIRST. See text.*

The largest GRG found by us is J1301+5105 in Figure 5. Its angular size of 10.3' and $z_{ph}$=0.275 (an average of values from 5 references) gives an LLS of 2.6 Mpc. Only a faint, diffuse core emission and the SE lobe is detected in NVSS, where it is unrecognizable as a GRG, demonstrating the superb sensitivity of LOFAR for this kind of objects.

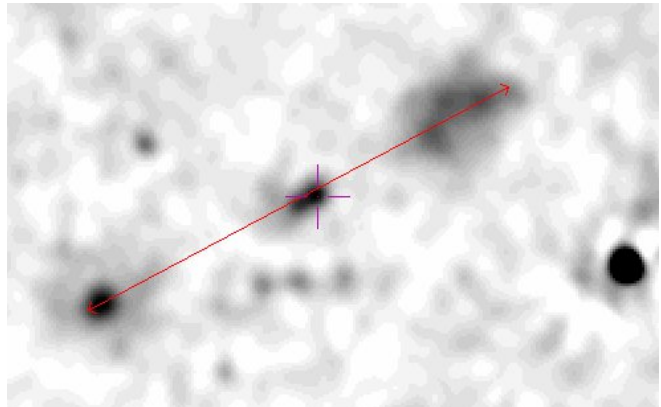

*Figure 5. GRG J1301+5105 in LoTSS-PDR. The cross marks the host and the straight line is 10.3' long.*

Many more candidates found in the present project remain to be inspected in more detail and will likely reveal more GRGs, especially radio-faint and distant ones. Note that while most of the ALR values of the GRGs found in this work are less than unity, the ALR for 240 GRGs with SDSS spectra has a median of ~0.9 and varies from 0.3 to 3.5, thus relativistic effects (like time delay, Doppler boosting) seem to have only a mild effect on source asymmetry.

## CONCLUSIONS

The present work, despite very limited in time to only five weeks, confirmed that visual inspection of radio images is a successful method for finding giant radio galaxies. Our results show that current and forthcoming low-frequency surveys with excellent sensitivity to low surface brightness features have a large potential to discover significant amounts of giant radio galaxies as well as sources of complex or currently unknown types of morphologies. Optical identification of GRGs relies heavily on the simultaneous availability of high-resolution surveys to locate the hosts, as well as on deep optical surveys to not only reveal the host but also to provide photometric redshift estimates. Automated algorithms to detect complex extended radio galaxies in survey images remain a challenge, but will be paramount to cope with the amount of data that will soon become available from the precursors of the Square Kilometre Array.

## ACKNOWLEDGEMENTS

We thank Tim Shimwell for access to the full atlas images of LoTSS-PDR. Brissa Gómez M., Braulio Arredondo P., and Douglas Monjardin W. helped inspecting some of the survey images. Josep Masqué kindly prepared the right panel of Figure 4, Roger Coziol provided useful comments, and Isabel Valdés O. improved the layout of this paper. H.A. benefitted from grant DAIP 980/2016-2017 of University of Guanajuato.